\documentstyle[prd,aps,floats,graphicx]{revtex}

\begin{document}
\preprint{hep-ph/0201209}
\draft
\input epsf
\renewcommand{\topfraction}{0.99}
\twocolumn[\hsize\textwidth\columnwidth\hsize\csname
@twocolumnfalse\endcsname

\title{Dynamics of a large extra dimension inspired hybrid inflation model} 
\author{Anne M Green~$^{1}$ and Anupam Mazumdar~$^{2}$} 
\address{$^{1}$ Astronomy Unit, School of Mathematical Sciences, 
Queen Mary University of London,\\ Mile End Road, London E1~4NS, 
United Kingdom\\ and \\ Physics Department, Stockholm University, 
S-106 91, Stockholm, Sweden (present address) \\
$^{2}$ The Abdus Salam International Centre for Theoretical Physics,
I-34100, Trieste, Italy\\} \date{\today} \maketitle
\begin{abstract}
In low scale quantum gravity scenarios the fundamental scale of nature
can be as low as TeV, in order to address the naturalness of the
electroweak scale. A number of difficulties arise in constructing
specific models; stabilisation of the radius of the extra dimensions,
avoidance of overproduction of Kaluza Klein modes, achieving
successful baryogenesis and production of a close to scale-invariant
spectrum of density perturbations with the correct amplitude. We
examine in detail the dynamics, including radion stabilisation, of a
hybrid inflation model that has been proposed in order to address
these difficulties, where the inflaton is a gauge singlet residing in
the bulk. We find that for a low fundamental scale the phase
transition, which in standard four dimensional hybrid models usually
ends inflation, is slow and there is second phase of inflation lasting
for a large number of $e$-foldings. The density perturbations on
cosmologically interesting scales exit the Hubble radius during this
second phase of inflation, and we find that their amplitude is far
smaller than is required.  We find that the duration of the second
phase of inflation can be short, so that cosmologically interesting
scales exit the Hubble radius prior to the phase transition, and the
density perturbations have the correct amplitude, only if the
fundamental scale takes an intermediate value.  Finally we comment
briefly on the implications of an intermediate fundamental scale for
the production of primordial black holes and baryogenesis.

\end{abstract}
\pacs{PACS numbers: 11.10Kk, 12.90+b, 98.80Cq}
\vskip2pc]

\section{Introduction}

In nature there are two apparent scales; the electroweak scale and the
scale of gravity, separated by seventeen orders of magnitude.
Understanding the gap between these scales has been a prime motivation
behind studying theories beyond the electroweak Standard Model (SM).
Supersymmetry provides an elegant scheme for keeping the electroweak
scale stable under any large radiative corrections, however the lack
of direct evidence for supersymmetry in collider physics and in nature
has lead to the consideration of scenarios with large extra
dimensions. In these scenarios the fundamental scale is taken to be
the higher dimensional Planck mass, $M_{\ast}$, which is assumed to be
close to the electroweak scale~\cite{nima0,early}. While in this
scheme supersymmetry is redundant in four dimensions, the presence of
low energy supersymmetry could still be a viable option, however, with
the fundamental scale at an intermediate scale, somewhere between the
Planck and electroweak scales. Such a scenario is well motivated by
string theory~\cite{quevedo}, which predicts that gauge and gravity
unification occurs below the Grand Unification Scale $\sim
10^{16.5}$GeV.

The four dimensional Planck mass in these theories is obtained
via dimensional reduction, assuming that the extra dimensions are
compactified on a torus, the simplest possible manifold. The volume of
the extra dimensions $V_d$, the effective four dimensional Planck
mass and the fundamental scale are then simply related:
\begin{equation} 
\label{imp}
M_{\rm p}^2 = M_*^{2+d} V_d \,,
\end{equation}
where $d$ is the number of extra compact dimensions. For given
$M_{\ast}$ this fixes the present day size of each of the extra
dimensions, $b_0$. For two extra dimensions and $M_{\ast} \sim 1$ TeV,
$b_0 \sim 0.2$ mm. Currently collider physics and supernovae $1987A$
impose a bound on the fundamental scale: $M_{\ast}\geq 30$ TeV
\cite{nima0,exp1,exp2}, and a recent astrophysical bound based on a heating
of a neutron star suggets $M_{\ast} \geq 500$~TeV \cite{new}.  
If the fundamental scale is as low as $\sim 100$ TeV, it is important 
that the SM particles are trapped in a four dimensional hypersurface 
(a $3$-brane) and are not allowed to propagate in the bulk~\cite{nima0}. 
It is generically assumed that besides gravity, the SM singlets, which 
may include the inflaton, can propagate in the bulk \cite{abdel100}.

The cosmological setup in models with large extra dimensions is quite
different from the conventional one. Firstly, if the electroweak scale
is the fundamental scale in higher dimensions, then there can be no
massive fields beyond the electroweak scale in four
dimensions. Secondly, the size of the extra dimensions can be quite
large, compared to the electroweak scale, which implies the existence
of new degrees of freedom, usually known as the radion, with a mass
scale as small as ${\cal O}(0.01 {\rm eV})$ if there are two large
extra dimensions. The large extra dimensions must grow from their
natural scale of compactification, $\sim(\rm TeV)^{-1}$, and then
stabilize at around a millimeter. This stabilization must occur before
the electroweak phase transition and nucleosynthesis, via some kind of
a trapping mechanism as discussed in Ref.~\cite{abdel2}. The Kaluza
Klein (KK) states of the graviton, and any other fields residing in
the bulk, can be excited at high temperatures and hence lead to
constraints on these models. Above the {\em normalcy temperature}, the
Universe could be filled by the KK modes. For Big Bang Nucleosynthesis
to occur successfully the normalcy temperature must be greater than
$\sim 1$ MeV. Furthermore the final reheat temperature, which is
constrained by cosmological considerations to be as small as $100$
MeV~\cite{nima0,davidson0,aemp}, should be smaller than the normalcy
temperature. These considerations severely restrict baryogenesis in
these models, for a detailed discussion see Refs.~\cite{abdel3,aemp}.

Constructing a successful inflation model, which produces a close to
scale invariant spectrum of density perturbations with the correct
amplitude and a very low reheat temperature is a challenging issue,
with single field models and models where the inflaton is a brane
field proving particularly problematic~\cite{many,anu}. There
have been several proposals~\cite{many,anu}, arguably the most natural
of which invokes SM singlet scalars coupled together to form a
potential which mimics that of the standard four dimensional hybrid
inflationary model, but with the fields promoted to the higher
dimensions~\cite{abdel1,abdel2}. It has also been shown that
baryogenesis can occur successfully in this model~\cite{abdel3,aemp}.
In this paper we study the dynamics of this extra dimension inspired
hybrid inflationary model in detail.

In hybrid inflation models, the false vacuum field is initially
trapped in a stable minimum at zero whilst the inflaton field
slow-rolls down its potential. At some critical value of the inflaton
field the stable minimum becomes an unstable maximum and quantum
diffusion produces a second order phase transition from the false
vacuum to the true vacuum~\cite{linde}. In standard four dimensional
cosmology, for most parameter values, the bare mass of the false
vacuum field is much greater than the Hubble parameter and the phase
transition occurs rapidly and inflation ends. If these quantities have
roughly the same magnitude, however, then the roll-down of the false
vacuum field is no longer fast and a second period of inflation
occurs~\cite{randall,garcia}. In standard four dimensional cosmology,
for the phase transition to occur slowly the effective coupling of the
false vacuum field has to be tiny, $\sim 10^{-30}$~\cite{garcia}. We
find, however, that for the parameter values which are relevant for
the extra dimensional model (fundamental coupling constants of order
unity and fundamental scale $\sim 100$TeV~\cite{abdel1,abdel2}), the
bare mass of the false vacuum field is of order the Hubble parameter
so that the phase transition is slow and a second period of inflation
occurs.

The duration of this second phase of inflation is typically very long,
so that the density perturbations on cosmological scales are generated
close its end. We calculate the amplitude of these perturbations and
find that they are only compatible with the COBE normalization if the
fundamental scale is significantly larger than 100 TeV. Finally we
discuss the implications, specifically achieving successfully
baryogenesis and avoiding the over-production of primordial black
holes (PBHs), of an intermediate fundamental scale. In order to keep
our discussion as general as possible, we do not fix either the
fundamental scale or the number of extra dimensions from the outset,
however we will focus throughout on the parameter values of the
specific model proposed in Refs.~\cite{abdel1,abdel2}.


\section{The model and its dynamics}

A single field inflationary model, either in four dimensions or with
the inflaton promoted into the bulk, can not provide adequate density
perturbations~\cite{abdel1}. This has led to the suggestion of a
hybrid inflationary model in higher dimensions with
potential~\cite{abdel1}:
\begin{eqnarray} 
V(\hat N,\hat\phi) = {\lambda^2 M_*^d}
\left(N_0^2 - {1\over M_*^d} \hat N^2\right)^2 + {m^2_{\phi}\over 2}
\hat\phi^2 \nonumber \\ + {g^2\over M_*^d} \hat N^2 \hat\phi^2 \,,
\label{hybp} 
\end{eqnarray} 
where $\hat\phi$ is the inflaton field and $\hat N$ is the subsidiary
false vacuum field which is responsible for the phase transition.  The
coupling constants, $g$ and $\lambda$ need not be identical, but we
assume them to be of ${\cal O}(1)$. The four dimensional Higgs vacuum
expectation value is determined by $ \lambda N_0$, and should be of
order $\sim {\cal O}(100)$ GeV. The higher dimensional field has a
mass of dimension $1+d/2$, which leads to non-renormalizable
interaction terms. The suppression, however, is given by the
fundamental scale, instead of the four dimensional Planck mass. Upon
dimensional reduction the effective four dimensional fields, $\phi$
and $N$, are related to their higher dimensional cousins by a simple
scaling
\begin{equation}
\label{rel}
 \phi =  \sqrt{V_d} \hat\phi \,, \quad \quad N =\sqrt{V_d} \hat N\,.
\end{equation}

From the point of view of four dimensions the extra dimensions are assumed 
to be compactified on a $d$ dimensional Ricci flat manifold with radii 
$b(t)$, with a minimum at $b_0$. The higher dimensional metric then reads
\begin{eqnarray}
 ds^2 = dt^2 -a^2(t)  d\vec x^2 -b^2(t)d\vec y^2\,,
\end{eqnarray}
where $\vec x$ denotes the three spatial dimensions, and $\vec y$ collectively
denotes the extra dimensions. The scale factor of the four dimensional 
space-time is denoted by $a(t)$. After dimensional reduction the effective 
four dimensional action reads 
\begin{eqnarray}
\label{action}
S=\int d^4x \sqrt{-g}\left[ -\frac{M_{\rm p}^2}{16\pi}R + \frac{1}{2}
\partial_{\mu} \sigma \partial^{\mu}\sigma -U(\sigma) \right. \nonumber \\
\left.  +\frac{1}{2}\partial_\mu \phi\partial^\mu \phi  +
\frac{1}{2}\partial_\mu N\partial^\mu N  
-\exp(-d\sigma/\sigma_0)V(\phi, N) \right]\,,
\end{eqnarray}
where the potential 
$V(\phi, N)$ can be derived from Eqs.~(\ref{hybp}) and (\ref{rel}):
\begin{eqnarray}
\label{hybp0}
V(\phi, N) \equiv \left(\frac{M_{\rm p}}{M_{*}}\right)^2\lambda^2 N_0^4 
+\frac{\lambda^2}{4}\left(\frac{M_{\ast}}{M_{\rm p}}\right)^2N^4 \nonumber \\
-\lambda^2 N_0^2N^2
+g^2\left(\frac{M_{\ast}}{M_{\rm p}}\right)^2\phi^2N^2 +\frac{1}
{2}m_{\phi}^2\phi^2\,,
\end{eqnarray}
and has a global minima at
\begin{equation}
\label{min}
\phi =0\,, \quad \quad N^2= 2 \left(\frac{M_{\rm p}}{M_{\ast}}\right)^2
N_{0}^2\,.
\end{equation}
In Refs.~\cite{abdel1,abdel2} the parameter values $N_{0}= M_{\ast}\sim 10^5$ 
GeV, $m_{\phi} \sim 10$ GeV and $\lambda \sim g \sim 1$ were taken.

The radion field $\sigma(t)$ can be written in terms of the radii of 
the extra dimension
\begin{eqnarray}
\label{radion0} 
\sigma(t) = \sigma_0 \ln \left[ \frac{b(t)}{b_0}\right]\,, \quad
\sigma_0 =
\left[\frac{d(d+2)M_{\rm p}^2}{16 \pi}\right]^{1/2} \,.
\end{eqnarray}
Note that $\sigma_0$ is proportional to the four dimensional Planck
mass. For illustrative purposes if we take the fundamental scale $\sim
{\cal O}(\rm TeV)$, the natural size of our three spatial dimensions,
and also the size of the extra spatial dimensions, is determined by
the fundamental scale $a(t)\sim b(t) \sim ({\rm TeV})^{-1}$.  From
Eq.~(\ref{imp}), assuming that there are only two extra dimensions,
the present size of the extra dimensions must be of order $b_0 \sim
{\cal O}(1 {\rm mm})$. The size of $b(t)$ must therefore expand from
$(\rm TeV)^{-1}$ and be stabilized at a mm i.e. there must be some
mechanism which traps the radion field in the minimum of the radion
potential, $U(\sigma)$. There is no concrete origin for this
potential, however the simplest possibility which gives the correct
mass for the radion is $U(\sigma)\sim m_{r}^2 \sigma^2$, where $m_{\rm
r} \sim 10^{-2}{\rm eV}$ for two extra spatial dimensions. A mechanism
which can trap the radion field in its potential was provided in
Ref.~\cite{abdel2} and we will now discuss its dynamics in detail.

Initially the dynamics of the universe are dominated by the
exponential potential of the radion. If $g N_0 \gg m_{\phi}$, the
false vacuum term dominates Eq.~(\ref{hybp0}): $ V(\phi, N) \approx
\lambda^2 (M_{\rm p}/M_{\ast})^2  N_{0}^{4} \approx const$. The 
exponential term, due to the radion field, multiplying the constant
scalar potential in Eq.~(\ref{action}) leads to a period
of power law asymmetric expansion for $a(t)$ and $b(t)$
\cite{berkin,anu,abdel2}:  
\begin{eqnarray}
\label{sol}
a(t) \sim {t}^{(d+2)/d}\,, \quad \quad \quad b(t) \sim {t}^{2/d}\,.
\end{eqnarray}
From the four dimensional point of view the radion drives a period of
inflation as it rolls down the exponential potential. Once it reaches
the critical value $|\sigma_0|$ (see Eq.~(\ref{action})) the effective 
mass of the radion field becomes of order the Hubble parameter
\begin{equation}
\label{mass}
m^2_{\rm r, eff} \approx m_{\rm r}^2+ \frac{V(\phi,N)}{\sigma_0^2} \sim 
{\cal O}(1)H^2\,,
\end{equation}
where we have neglected the contribution from $U(\sigma)$ as it is
small compared to that from $V(\phi,N)$ ($U(\sigma) \sim M_{\rm p}^2
m_{\rm r}^2 \ll V(\phi,N) \sim M_{\rm p}^2 M_{\ast}^2$, as $\sigma(t)
\rightarrow \sigma_0 \approx M_{\rm p}$).  At this point the radion
field can no longer support inflation, however inflation continues as
the $\phi$ field slowly rolls down the potential, with Hubble parameter
\begin{equation}
\label{hub0}
H \approx \sqrt{\frac{8\pi}{3}}\frac{\lambda N_0^2}{M_{\ast}}\,.
\end{equation}
Subsequently the mass of the radion field is dominated by the
Hubble constant, $H$, and the radion field approaches 
the global minimum configuration, $\sigma=0$, exponentially fast: 
\begin{equation} 
\label{rest}
\sigma(t) \approx \sigma_0 e^{\left( - {m_{\rm r, eff}^2(H) t/ 3H} \right)} 
\sim \sigma_0 e^{\left(- {H t / 3}\right)} \,, 
\end{equation}
so that the radion is trapped in its own potential $U(\sigma)$, within
a Hubble time, and the radion configuration remains dynamically frozen
while inflation continues. As the radion couples to the trace of the
energy momentum tensor it therefore also couples to the SM particles,
which are essentially the decay products of $\phi$ and $N$, and the
Hubble induced correction remains even after the end of inflation.
Note that the radion interaction with the SM fields is extremely weak,
due to the Planck mass suppressed couplings, and hence it can never
reach thermal equilibrium. At a certain energy scale, when the bare
mass of radion mass comes to dominate the Hubble induced correction,
the radion begins to oscillate around the minimum of the potential
with an amplitude $\propto m_{\rm r}$. If the bare mass of the radion
is very small, $\lesssim 10^{-2}$eV in the case of two large extra
dimensions and the fundamental scale or order a TeV, then the radion
density stored in the oscillations is not large enough to act as
dominant component of the total energy density of the Universe.
However, for a mass as small as $10^{-2}$eV, the oscillations are
rapid, $\nu \sim m_{\rm r}^{-1}\sim 10^{11}$Hz, which suggests that
Newtons constant may get a time varying contribution from the radion
oscillations~\cite{steinhardt}, so that $\dot G/G\neq 0$. In theories
where Newtons constant is time varying, such as the Brans-Dicke theory of
gravity \cite{bd}, the time variation in $G$ contributes to the total
energy density and therefore affects the expansion rate of the
Universe. Even though, the amplitude of the radion oscillations is
small, the high frequency of the oscillations may provide a
significant contribution to the Hubble expansion. This is an
interesting topic which merits a separate study.

We will now concentrate on the subsequent dynamics of the $\phi$ and
$N$ fields. The $N$-field is rapidly driven to the false vacuum, $N
=0$, and the $\phi$ field rolls slowly down the potential until it
reaches the critical value, $\phi_{{\rm c}}$,
\begin{equation}
\label{critical} 
\phi_{\rm c} = \frac{\lambda}{g}\left(\frac{N_0 M_{\rm p}}{M_{\ast}}\right)\,,
\end{equation}
where the effective mass squared of the $N$ field becomes negative and
a second order phase transition begins.  Note that if $\lambda \sim g
$, and, $N_0 \sim M_{\ast}$, then $\phi_{\rm c} \sim M_{\rm p}$. The
effective mass of the false vacuum field, $m_{{ N}}$, is
\begin{equation}
m_{{N}}= \sqrt{2} \lambda N_{0} \,,
\end{equation} 
so that
\begin{equation}
\frac{m_{{ N}}}{H} \approx \sqrt{\frac{3}{4 \pi}}\frac{M_{\ast}}{N_{0}} \,.
\end{equation}
For $M_{\ast} \sim N_{0}$, $H \sim m_{{ N}}$, so that the phase
transition occurs slowly and there is a second period of inflation as
the fields initially roll slowly towards the global minimum of the
potential.

The dynamics of hybrid inflation models where there is a second period
of inflation have previously been studied by Randall,
Solja\u{c}i\'{c}, and Guth~\cite{randall} and Garc\'{\i}a-Bellido,
Linde, and Wands\cite{garcia} (GLW hereafter), with particular focus on
the production, at the phase transition, of large density
perturbations, which may lead to the over-production of primordial
black holes. GLW parameterise the potential as
\begin{equation}
\label{glw} 
V(\phi,\psi) = \left( M^2 - \frac{\sqrt{\tilde{\lambda}}}{2} \psi^2 \right)^2 +
    \frac{1}{2} \tilde{m}_{\phi}^2 \tilde{\phi}^2 + \frac{1}{2} \gamma
    \tilde{\phi}^2 \psi^2 \,\,
\end{equation}
and find that $m_{\phi} \sim H \sim 1$TeV if $M \sim 10^{11}$GeV, $m
\sim 10^3$ GeV and $\lambda \sim 10^{-30}$. Comparing Eqs.~(\ref{hybp0})
and (\ref{glw}) one finds the following parameter equivalences: $ \phi \equiv
\tilde{\phi}   \,, m_{\phi}  \equiv  \tilde{m}_{\phi} \,, $
\begin{eqnarray}
 \frac{\lambda}{2} \left( \frac{M_{\rm P}}{M_{\ast}} \right) N_{0}^2 
  & \equiv & M^2 \,, \nonumber   \\
  \frac{g^2}{\lambda} \left( \frac{M_{\ast}}{M_{\rm P}} \right) 
  & \equiv & \frac{1}{2} \frac{\gamma}{\sqrt{\tilde{\lambda}}}  
       \,, \nonumber  \\
      \lambda \left( \frac{M_{\ast}}{M_{\rm P}} \right) N^2 & \equiv & 
        \sqrt{ \tilde{\lambda} }  \psi^2 \,.
\end{eqnarray}
So we see that if $M_{\ast} \sim N_0 \sim 10^5$ GeV then we
automatically obtain the extremely small effective coupling for the
false vacuum field, required for a slow phase transition and second
period of inflation, for values of the fundamental couplings $g$ and
$\lambda$ of order unity. One important difference between our model
and that of GLW is the critical value of the slow-rolling field. GLW
have
\begin{equation}
\tilde{\phi_{{\rm c}}}^2  = \frac{2 \sqrt{\lambda}}{\gamma} M^2 \sim 
         \frac{10^{-31}}{\gamma} M_{\rm P}^2 \,,
\end{equation} 
whereas we have $\phi_{{\rm c}} \sim M_{\rm P}$. This suggests that in our
case the dynamical behavior of the fields is quite different from that
of previously studied hybrid inflation models, even those where a slow
phase transition occurs~\cite{randall,garcia}.

We will now study the dynamics of the fields for our parameters
analytically. Once the radion has stabilized and the false
vacuum field $N$ has evolved to $N=0$, provided that $ g N_0 \gg
m_{\phi}$ the false vacuum term dominates the potential, so that the
evolution of the $\phi$ field is given by:
\begin{equation}
\phi= \phi_{\rm i} \exp{ \left( - \frac{1}{\sqrt{6 \pi} \lambda}
         \frac{ M_{\ast} m_{\phi}^2}{N_{0}^2} t \right)} \,,
\end{equation}
where $\phi_{\rm i}$ is the initial value at some arbitrary initial time $t=0$.
The number of e-foldings of inflation which occur between
$\phi_{\rm i}$ and $\phi$ is given by:
\begin{equation}
N_{\rm e} = 2 \pi \lambda^2 \frac{N_{0}^4}{M_{\ast}^2 m_{\phi}^2} \ln{
           \frac{ \phi_{\rm i}}{\phi}} \,.
\end{equation}
We see that for $N_{0} \sim M_{\ast} \gg m_{\phi}$ the evolution of the
$\phi$ field is extremely slow, and the duration of this first phase
of inflation is large.

The evolution of $N$ as $\phi \rightarrow \phi_{\rm c}$ is more
complicated, as $m_{N}^2 < H^2$ once
\begin{equation}
\phi^2 < \phi_{{\rm c}}^2 + \frac{ 4 \pi}{3} \left( \frac{\lambda N_{0}^2
         M_{\rm P}}{M_{\ast}} \right)^2 \,,
\end{equation}
so that the quantum fluctuations in the $N$ field become important.
The Fokker-Planck equation~\cite{fp} is usually employed to study the
dynamics of the $N$ field in this regime~\cite{randall,garcia}, using
the assumptions that the field has a delta-function distribution at
some initial time (when $\phi \gg \phi_{{\rm c}}$) and that the
average quantum diffusion per Hubble volume per Hubble time is
$\approx H/2 \pi$.  It was found in Ref.~\cite{garcia} that the
typical value of the $N$ field when $\phi=\phi_{\rm c}$, $\bar{N}$, is
given by
\begin{equation}
\label{diffu}
\bar{N}^2= \left( \frac{H}{2 \pi^2} \right)^2 \frac{1}{2 r} \left( \frac{
           e^a}{a}\right)^a \Gamma(a,a) \,,
\end{equation}
where $a= (4 \lambda^2 N_{0}^2 / 3 m_{\phi}^2)$, $\Gamma(a,a)$ is the
Incomplete Gamma function and
\begin{equation}
r  = \frac{3}{2}  - \sqrt{ \frac{9}{4} - \frac{m_{\phi}^2}{H^2}} \approx
             \frac{m_{\phi}^2}{3 H^2} \,.
\end{equation}     
Eq.~(\ref{diffu}) must be evaluated numerically; for example if
$\lambda=g=1$, $N_{0}= M_{\ast}= 10^5$ GeV and $m_{\phi}=10$ GeV, as
in Refs.~\cite{abdel1,abdel2}, then $\bar{N} = 22 (H/ 2 \pi)$.

In order to study the second phase of inflation which occurs after
$\phi=\phi_{\rm c}$, it will prove useful to rewrite Eq.~(\ref{hybp0})
using Eq.~(\ref{critical}) as
\begin{eqnarray}
V(\phi, N) \equiv \left(\frac{M_{\rm p}}{M_{*}}\right)^2\lambda^2
N_0^4 +\frac{\lambda^2}{4}\left(\frac{M_{\ast}}{M_{\rm p}}\right)^2N^4
\nonumber \\ +g^2\left(\frac{M_{\ast}}{M_{\rm p}}\right)^2 N^2
(\phi^2-\phi_{{\rm c}}^2) +\frac{1}{2}m_{\phi}^2\phi^2\,,
\end{eqnarray}      
Furthermore if $N_{0} \gg m_{\phi}$ then for $N \ll (M_{\rm P}
N_{0}/M_{\ast})$ the false vacuum term in the potential dominates and
the Hubble parameter remains constant.

The slope of the potential in the $\phi$ direction is given by
\begin{equation}
\label{deriv1}
\frac{{\rm d}V}{{\rm d} \phi} = \left[ 2 g^{2} \left( \frac{M_{*}}{M_{\rm P}} 
\right)^{2} N^{2} + m_{\phi}^{2} \right] \phi \,.
\end{equation}
Provided that $(10 \lambda N_0^2) / (g m_{\phi} M_{\rm P}) \ll 1$ then the
second term dominates initially and $\phi$ evolves away
from $\phi_{\rm c}$ with equation of motion
\begin{equation}
\label{phi2}
\phi=\phi_{c} \exp{ \left[ - \frac{1}{\sqrt{24 \pi}} \frac{1}{\lambda}
        \left( \frac{m_{\phi}}{N_{0}} \right)^{2} (M_{\ast} \hat{t})
        \right]} \,,
\end{equation}
where we have now taken $\hat{t}=0$ when $\phi=\phi_{\rm c}$.

In the $N$ direction the slope of the potential is given by
\begin{equation}
\label{deriv2}
\frac{{\rm d}V}{{\rm d} N} = \left[ \lambda ^{2} N^{2} + 2 g^2 (\phi^{2} 
-\phi_{c}^{2} ) \right] N \left( \frac{M_{*}}{M_{\rm P}} \right)^{2} \,.
\end{equation}
If $\bar{N}$ is sufficiently large (as is the case for the specific
parameter values we are interested in: $\bar{N}= 22 (H/ 2 \pi)$) then
there is a small period where the first term dominates and $N(\hat{t})
\sim \bar{N}$. As $\phi$ evolves away from $\phi_{{\rm c}}$, however,
the second term soon comes to dominate, so that for small $\hat{t}$,
using the first order expansion of Eq.~(\ref{phi2})
\begin{equation}
\phi-\phi_{c} \sim - \frac{1}{\sqrt{24 \pi}} \frac{1}{g} \left( 
\frac{m_{\phi}^2}{N_{0} }\right)M_{\rm p} t \,,
\end{equation}
the $N$ field grows exponentially
\begin{equation}
\label{n2}
N= \bar{N} \, \exp{ \left[ \frac{1}{12 \pi} \left( \frac{m_{\phi}}{N_{0}} 
\right)^{2} \left( M_{\ast} t \right)^{2} \right]}  \,,
\end{equation}
where we have neglected the initial period where $N \sim \bar{N}$
since its duration is negligible compared with that of the subsequent
exponential growth.  For the majority of the duration of the second
phase of inflation the $N$ field grows exponentially and $\phi$ moves
slowly away from $\phi_{{\rm c}}$.  Once
\begin{equation}
\label{end}
2 g^{2} \left( \frac{M_{*}}{M_{\rm P}} \right)^{2} N^{2} \sim m_{\phi}^{2} \,,
\end{equation}
however, the first term in Eq.~(\ref{deriv1}), which is growing
exponentially, comes to dominate the evolution of the $\phi$ field and
causes the $\phi$ field to evolve rapidly away from $\phi_{\rm c}$. At
this point $N \sim (\phi_{{\rm c}} -\phi)$ so that ${\rm d}V/{\rm d}N
\sim {\rm d}V/{\rm d}\phi$ and both $N$ and $(\phi_{{\rm c}} -\phi)$
grow rapidly, and inflation comes to an end shortly afterwards, with
both fields subsequently oscillating about the global minimum of the
potential. We can therefore use the time at which Eq.~(\ref{end}) is
satisfied to estimate the duration of the second phase of inflation,
$t_2$:

\begin{equation}
\label{time}
t_{\rm 2} \approx \frac{\sqrt{6 \pi}  N_{0}}{m_{\phi} M_{\ast}} 
\left\{\ln\left[ \frac{1}{2 g^2} \left(\frac{M_P}{M_{\ast}}\right)^2\left
(\frac{m_{\phi}}{\bar N}\right)^2\right]\right\}^{1/2}\,.
\end{equation}
Since the Hubble parameter remains constant until very close to the end
of the second phase of inflation, we can estimate the total number of
e-foldings which occur during the second phase of inflation as:
\begin{eqnarray}
\label{ne2}
N_{\rm e2} & \approx & H t_{2} = 4 \pi \lambda
\frac{N_{0}^3}{M_{\ast}^2 m_{\phi}} \nonumber \\
&&  \times \left\{\ln\left[
2\sqrt{\pi}\left(\frac{N_0}{M_{\ast}}\right)^2\left(\frac{M_{\rm p}}{\bar N}
\right)\right]\right\}^{1/2}\,.
\end{eqnarray}
For the parameter values used in Refs.~\cite{abdel1,abdel2} ($N_0 \sim
M_{\ast} \sim 10^{5}$GeV, $m_{\phi} \sim 10$GeV, and $\lambda \sim g
\sim {\cal O}(1)$) we obtain $N_{\rm e2} \sim 9\times 10^{5}$, an
extremely large number of e-foldings of inflation. This reiterates the
point that for $N_0 \sim M_{\ast}$ the phase transition is extremely
slow.

This description is borne out by numerical evolution of the full
equations of motion of the fields. We find that, for the set of
parameter values above, the second phase of inflation lasts for $8.5
\times 10^5$ $e$-foldings, and that $\phi \sim \phi_{{\rm c}}$ and $N
\sim 0$ until the last 100 $e$-foldings or so. In Fig. 1 we plot the
evolution of the fields during the late stages of the second phase of
inflation. Note that the evolution of the fields away from
$\phi=\phi_{\rm c}$ and $N=0$ at the end of the second phase of
inflation is so rapid that if we plotted their evolution for the
entire second period of inflation linearly, only the straight lines
$\phi=\phi_{\rm c}$ and $N=0$ would be visible.

\begin{figure}
\centering
\leavevmode\epsfysize=6.5cm \epsfbox{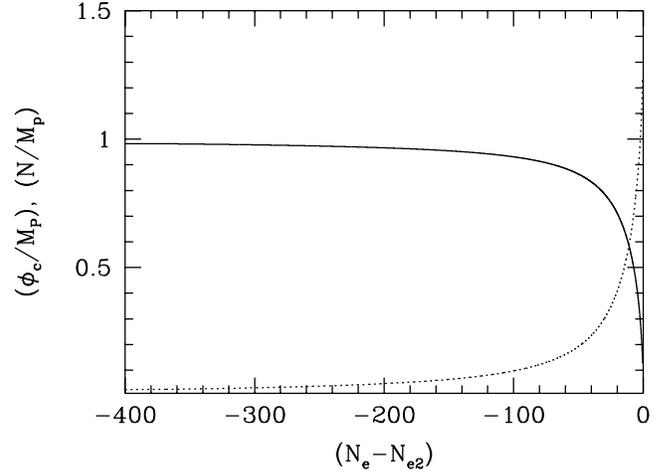}\\
\caption[fig1]{\label{fig1} The evolution of the $\phi$ (solid) and
$N$ (dotted) fields as a function of minus the number of $e$-foldings
from the end of inflation (so that time flows from left to right),
towards the end of the second phase of inflation, for the parameter
values $N_0 = M_{\ast} = 10^{5}$GeV, $m_{\phi} = 10$GeV, and
$\lambda = g = {\cal O}(1)$, so that $N_{{\rm e2}} \approx 8.5
\times 10^5$.}
\end{figure}

\section{Density perturbations}

In this model the scales relevant for the microwave background and
large scale structure exit the Hubble radius $43$ $e$-foldings before the
end of inflation~\cite{abdel1}, this is less than the usual $60$
$e$-foldings because of the small inflationary energy scale and the
requirement that the reheat temperature should be $\sim {\cal
O}(10-100)$MeV, in order not to overproduce bulk matter such as
excitation of KK modes, for details see Ref.~\cite{aemp}.  The
amplitude of the density perturbations on scales which leave the
Hubble radius towards the end of the first period of inflation, when $\phi
\sim \phi_{\rm c}$ and $N \sim 0$, can be calculated
easily~\cite{abdel1,aemp}:
\begin{equation}
\label{imp1}
\delta_{\rm H} =8.2\lambda^2 g \frac{N_0^5}{M_{\ast}^2 m_{\phi}^2 M_{\rm p}}\,.
\end{equation}
For $\lambda \sim g \sim 1$, $N_{0} \sim M_{\ast} \sim 10^5$ GeV then
to produce the COBE normalization, $\delta_{\rm H} \simeq 1.95 \times
10^{-5}$~\cite{bunn}, requires $m_{\phi} \sim 10$
GeV~\cite{abdel1,abdel2}, however we have already seen that for these
parameter values $N_{\rm e2} \sim 9 \times 10^5 \gg 43$, so that the
scales that left the Hubble radius during the first period of
inflation have yet to re-enter the Hubble radius. In other words the
scales which are cosmologically interesting leave the Hubble radius
during the second phase of inflation. Combining Eqs.~(\ref{ne2}) and
(\ref{imp1}) we find that requiring the perturbations on the scales
probed by COBE to leave the Hubble radius during the first phase of
inflation {\em and} have the correct amplitude ($N_{{\rm e2}} < 43$
and $\delta_{{\rm H}} =1.9 \times 10^{-5}$) requires
\begin{equation}
\frac{M_{\rm P} N_0}{g M_{\ast}^2} \left[ \ln{ \left\{ 2\sqrt{\pi}
             \left(\frac{N_0}{M_{\ast}}\right)^2\left(\frac{M_{\rm p}}{\bar N}
             \right) \right\} } \right] \lesssim 10^6 \,,
\end{equation} 
so for $M_{\ast} \sim 10^5$ GeV and $ g \sim {\cal O} (1)$ we would
need $N_{0} \lesssim 10^{-4}$ GeV i.e. if $M_{\ast} \sim 10^5$ GeV
then for $N_{0} > 10^{-4}$ GeV there is {\em no} value of $m_{\phi}$
for which $N_{{\rm e2}} < 43$ and $\delta_{{\rm H}} =1.9 \times
10^{-5}$.  The presence of such a small vacuum expectation value for
the $N$ field and a negligible bare mass for the $\phi$ field is an
extreme fine tuning in $4+d$ dimensions which is unlikely, so we will
not pursue this possibility further.

Later we will examine whether it is possible to construct a satisfactory 
model where the cosmologically interesting density perturbations are 
produced during the first phase of inflation and have the correct 
amplitude, by employing an intermediate fundamental scale, however for 
now we will continue to focus on the parameter values used in 
Refs.~\cite{abdel1,abdel2}.

For these parameters values the cosmologically interesting density
perturbations are produced during the second phase of inflation, when
both fields are dynamically important, and to calculate their
amplitude we therefore need to employ the formula for multiple
slow-rolling scalar fields~\cite{msr}:
\begin{equation}
\label{estim1}
\delta_{\rm H}^2  = \frac{1}{75 \pi^2} \left( \frac{8 \pi}{M_{\rm p}^2} 
                  \right)^3  V^3 \left[ \left( \frac{{\rm d}V}{{\rm d} \phi}
                   \right)^2 +  \left( \frac{{\rm d}V}{{\rm d} N}
                   \right)^2 \right]^{-1} \,.
\end{equation}
The scales that we are interested in leave the Hubble radius very
close to the end of the second period of inflation 
($43/(8.5 \times 10^5) \ll1$) when both fields are evolving 
rapidly and it is not possible to accurately follow their motion 
analytically. We therefore evolve the fields numerically and utilize 
Eq.~(\ref{estim1}) to evaluate $\delta_{\rm H}$. We can make several 
simple observations, however. After the beginning of the phase transition, 
but far from the end of inflation, 
${\rm d}V/{\rm d} \phi \approx m_{\phi}^2 \phi \gg {\rm d}V/{\rm d}N$ 
so that $\delta_{\rm H}$ has the same value as prior to the phase 
transition, given by Eq.~(\ref{imp1}). The fields begin evolving more 
rapidly during the last 100 or so $e$-foldings of inflation with 
${\rm d}V/{\rm d}N$ and ${\rm d}V/{\rm d}\phi$, which are of the same 
order of magnitude, increasing significantly, so that $\delta_{\rm H}$ 
decreases. We can make an order of magnitude estimate of $\delta_{\rm H}$ 
at the end of inflation, $\delta_{\rm H}(\epsilon=1)$, by pretending that 
only one of the fields is dynamically important and utilizing the 
single-field expression for $\delta_{\rm H}$ in terms of the first 
slow-roll parameter $\epsilon_{\phi} \equiv(M_P^2/ 16 \pi) (V'/V)^2$:
\begin{equation}
\delta_{\rm H} = \frac{32}{75 M_{\rm P}^4} \frac{V}{\epsilon} \,.
\end{equation}
Inflation ends when $\epsilon=1$, so that~\footnote{This is an upper-limit 
since we are neglecting the change in $V$.}
\begin{equation}
\label{delend}
\delta_{\rm H}(\epsilon=1) < \frac{\lambda N_0^2}{2 M_{\ast} M_{\rm P}}
\,.
\end{equation}
For the parameters used in Refs.~\cite{abdel1,abdel2} this gives
$\delta_{\rm H}(\epsilon=1) < 4 \times 10^{-15}$. The large change in
$\delta_{\rm H}$ towards the end of inflation is, along with the long
duration of the second phase of inflation, a consequence of the
extremely slow evolution of the fields at the beginning of the phase
transition.

\begin{figure}
\centering
\leavevmode\epsfysize=6.5cm \epsfbox{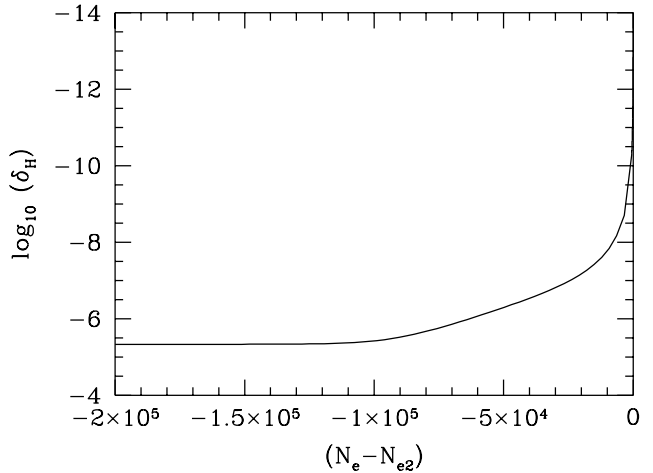}\\ 
\caption[fig2]{\label{fig2} The amplitude of the density
perturbations, $\delta_{\rm H}$, as a function of minus the number of
$e$-foldings before the end of inflation for $\lambda = 1$, 
$N_{0} = M_{\ast} \sim 10^5$ GeV and $m_{\phi} = 10$GeV.}
\end{figure}

The variation of $\delta_{\rm H}$ with number of $e$-foldings before
the end of inflation is shown in Fig. 2 . We find that 43-foldings
before the end of inflation $\delta_{\rm H} \sim 3 \times 10^{-13}$,
far smaller than required by the COBE normalization. We also find that
$\delta_{\rm H}$ at the end of inflation, is independent of
$m_{\phi}$, as expected from Eq.~(\ref{delend}). If $m_{\phi}$ is
decreased, with the other parameters kept fixed, while $\delta_{\rm
H}$ on scales which exit the Hubble radius well before the end of
inflation is increased, as predicted by Eq.~(\ref{estim1}), the value
43 $e$-foldings before the end remains stubbornly at $3\times
10^{-15}$. Therefore for the parameters of the model proposed in
Refs.~\cite{abdel1,abdel2,abdel3} it is not possible to produce
density perturbations of the required size.

We will now examine whether it is possible to choose parameter values
such that $N_{{\rm e2}} < 43$, so that the perturbations on the scales
probed by COBE exit the Hubble radius during the first period of
inflation, and $\delta_{\rm H}$, which is now given by
Eq.~(\ref{estim1}), satisfies the COBE normalization. This can be
achieved if the fundamental scale of gravity is at some intermediate
scale. For instance if $M_{\ast}\sim 10^9$ GeV, with $N_{0}\sim 10^4$
GeV, and $m_{\phi}\sim 10^{-5}$ GeV we find numerically that $N_{{\rm
e2}} < {\cal O }(10)$~\footnote{The assumptions employed in the
derivation of Eq.~(\ref{ne2}) break down for these parameter values.}
and $\delta_{\rm H} \sim {\cal O}(10^{-5})$. Note that even though
$m_{{\rm N}} > H$ for these parameter values, there is still a short
second phase of inflation due to the small gradients, and consequent
slow evolution, of the fields at the beginning of the phase
transition.  These parameters values should be considered as
illustrative, as they are clearly not unique.


\section{Implications of an intermediate fundamental scale}

We will now study some other implications of an intermediate
fundamental scale for the model.
\subsection{Primordial black hole production}

For the set of parameter values considered in
Refs.~\cite{abdel1,abdel2,abdel3} $N_0 \sim M_{\ast} \sim 10^{5}$ GeV,
$m_{\phi} \sim 10$GeV, and $\lambda \sim g \sim {\cal O}(1)$ the
duration of the second phase of inflation is so long that the density
perturbations on scales corresponding to the beginning of the phase
transition, which may be large and lead to PBH
overproduction~\cite{randall,garcia}, remain well outside the Hubble
radius today. For parameter values where there is a second phase of
inflation with duration less than 43 $e$-foldings, these potentially
large fluctuations will have re-entered the Hubble radius by the
present day and we need to worry about PBH overproduction. The
production of PBH in extra dimensional scenarios at colliders and from
high energy cosmic-rays has been studied in great detail
recently~\cite{pbhed}. PBHs produced in the early universe, via the
collapse of large density pertubarions, form via a completely
different mechanism however. The evaporation of cosmological BHs in
extra-dimensional scenarios has been examined~\cite{pbhedevap}, but
the conditions for PBH formation in these scenarios are not known with
any precision. We will therefore use the calculations of GLW to
examine the order of magnitude constraint on the parameters of our
model.

The amplitude of the fluctuations on scales corresponding to the
beginning of the phase transition can be estimated as~\cite{garcia}
\begin{equation}
\delta \equiv \frac{\delta \rho}{\rho} \approx \frac{4}{9s} \,,
\end{equation}
where
\begin{equation}
s = - \frac{3}{2} \pm \sqrt{ \frac{9}{4} + \frac{m_{{\rm N}}^2}{H^2}} \,.
\end{equation}
If $\delta \ll 1$ then the present day density of PBHs will be
negligible (see eg.~\cite{pbh}). This is guaranteed if
\begin{equation}
\label{cond}
\frac{m_{\rm N}}{H} = \sqrt{\frac{3}{4 \pi}}\frac{M_{\ast}}{N_{0}} \gg
1 \,.
\end{equation}
This is precisely the condition usually given for inflation to end
promptly at $\phi = \phi_{{\rm c}}$~\cite{garcia}, implying that a
slow phase transition automatically leads to the formation of a
non-negligible population of PBHs. We have seen, however, that if the
gradients of the fields are sufficiently small a second phase of
inflation, with non-negligible duration, can still occur even when
this condition is marginally satisfied. We therefore conclude that to
find whether PBHs are overproduced for the set of parameters
suggested above, would require a careful study of the formation of
PBHs in extra dimensional scenarios, and also an accurate calculation
of the amplitude of the density perturbations produced at the
beginning of the phase transition.

An attractive alternative would be to avoid the possibility of PBH
overproduction entirely, by choosing parameter values for which the
phase transition occurs rapidly and there is no second phase of
inflation. In the extra dimensional model which we are studying to
ensure this, while producing density perturbations of the required
magnitude on cosmological scales, would require an even larger
fundamental scale however.

\subsection{Baryogenesis}

For TeV scale quantum gravity with a normalcy temperature as low as
$\sim 10$MeV, the only calculable and predictable model of
baryogenesis has been given in Ref.~\cite{aemp,abdel3}. Other avenues
for baryogenesis exist in brane world scenarios where the Standard
Model fields also propagate in the bulk. In this case it is possible
to localise the fermionic wavefunction on a brane in such a way that a
small overlap prevents fast proton decay~\cite{schmaltz}. Baryogenesis
in these models has been explored in detail in Ref.~\cite{rula},
however the model which we are studying is quite different, as it is
assumed that the Standard Model fields are confined to the brane.

The general idea is based on the fact that there exists a gauge
singlet which carries a $U(1)$ charge, which is dynamically broken at
a scale governed by the Hubble expansion after the end of inflation
\cite{abdel3}. This introduces a broken $C$ and $CP$ phase. As a
result the real and imaginary parts of the $U(1)$ carrying field
spiral and produce an initially asymmetric distribution of field
quanta, which is transferred to the SM quarks and leptons via the
dimensional six baryon number violating lepto-quark operator. Note,
that such an operator can also induce proton decay as long as the
gauge singlet responsible for baryogenesis never develops any vacuum
expectation value.

The additional second phase of inflation which we have found will not
affect the dynamical mechanism of producing baryogenesis which was
provided in Ref.~\cite{aemp,abdel3}. However, in order to obtain the
observed baryon asymmetry $\sim 10^{-10}$, the gauge singlet must have
a large initial amplitude of order $\sim 10^{16}$~GeV. This could only
be obtained if the singlet were promoted to the bulk along with the
inflaton sector. Now, if we raise the fundamental scale to $\sim
10^{9}$ GeV, we certainly relax the stringent constraint on the
normalcy temperature for two extra dimensions from $100$MeV to $1$
TeV,~\cite{aemp}. In this case our Universe could afford to have a
reheat temperature of the order of electroweak scale, which opens up
the possibility of electroweak baryogenesis and leptogenesis. Note
that the baryogenesis mechanism which was described above very
briefly, and in detail in Ref.~\cite{aemp,abdel3}, would remain a
viable option. Here again one has to promote the $U(1)$ carrying gauge
singlet to the bulk along with the inflaton. This is because by
raising the fundamental scale we are also raising the energy density
stored in the inflaton sector, and in order to get the
right baryon to photon ratio one then has to have a larger amplitude,
comparable to the Planck scale now, for the gauge singlet.

\section{Conclusions}

We have studied in detail the dynamics of a hybrid inflationary model
in the context of large extra dimensions, proposed in
Ref.~\cite{abdel1} and subsequently studied in
Refs.~\cite{abdel2,abdel3,aemp}, where it is assumed that the inflaton
sector is a gauge singlet residing in the higher dimensions. We have
studied the entire gamut of the dynamical behavior of the coupled
scalar fields. In particular we have shown that for a low fundamental
scale, as studied in Refs.~\cite{abdel1,abdel2,abdel3}, there are in
fact two distinct phases of inflation.  The first phase of inflation
has two parts, an initial period of radion dominated inflation, as the
radion is stabilised via dynamical trapping in its own potential,
followed by vacuum dominated inflation, as in standard hybrid
models. At low energies, when the radion mass dominates the Hubble
parameter, the radion begins oscillating around the minimum of its
potential, however if the bare mass of the radion is very small,
$\lesssim 10^{-2}$eV in the case of two large extra dimensions and the
fundamental scale ${\cal O} ({\rm TeV})$, then the radion density
stored in the oscillations is not large enough to cause problems
similar to the moduli problem.

For the parameter values considered natural in standard four
dimensional hybrid inflation models, inflation ends rapidly, once the
$\phi$ field reaches the critical value at which the false vacuum
becomes unstable.  For the parameters which are natural from an extra
dimensional perspective (a fundamental scale $M_{\ast} \sim N_0 \sim
10^{5}$ GeV and fundamental coupling constants $\lambda,g \sim {\cal
O}(1)$) we find that the phase transition is slow, producing a second
phase of inflation which lasts for around $10^{6}$
$e$-foldings. Cosmologically interesting scales therefore exit the
Hubble radius close to the end of this second period of inflation, and
we find that the amplitude of the density perturbations on these
scales is smaller than required by the COBE normalisation unless the
vacuum expectation value of the false vacuum field, $N_{0}$, is less
than $10^{-4}$ GeV.

The only way around this obstacle is to try to shorten the second
phase of inflation so that cosmologically interesting scales exit the
Hubble radius before the phase transition. To do this we need a
fundamental scale higher than $10^{5}$GeV. We have found that, for
instance, with the set of parameter values $M_{\ast}\sim 10^9$GeV,
$N_{0}\sim 10^4$GeV and $m_{\phi}\sim 10^{-5}$GeV density
perturbations of the correct amplitude are produced. For this
intermediate fundamental scale the constraint on the normalcy
temperature is relaxed so that electroweak baryogenesis and
leptogenesis become possible, furthermore the baryogenesis mechanism
provided in Ref.~\cite{aemp} is unaffected by the second phase of
inflation and remains viable. If there is a second phase of inflation
lasting less than 43 $e$-foldings then the large density perturbations
produced at the beginning of the phase transition will have re-entered
the Hubble radius by the present day, and may lead to the
over-production of PBHs~\cite{randall,garcia}. The formation criteria
for PBHs in extra-dimensional scenarios have not yet been studied, so
it is not possible to use the constraints on PBH abundance~\cite{pbh}
to constrain the model parameters.

\section*{Acknowledgments}

A.M.G.~was supported by PPARC and the Swedish Research Council. We
thank Abdel P\'erez Lorenzana for useful comments, and Andrew Liddle
for numerous useful discussions in the early stages of this work. 
A.M.G.~acknowledges use of the Starlink computer system
at Queen Mary, University of London.


\begin{references}
\bibitem{nima0}
N.Arkani-Hamed, S. Dimopoulos, and G. R. Dvali,
Phys. Lett B {\bf 429}, 263 (1998); \prd {\bf 59}, 086004 (1999);
I. Antoniadis, N. Arkani-Hamed, S. Dimopoulos, and G. R. Dvali,
Phys. Lett. B {\bf 436}, 257 (1998).

\bibitem{early}
For some early ideas  see also:
I. Antoniadis, Phys. Lett. B {\bf 246}, 377 (1990);
I. Antoniadis, K. Benakli and M. Quir\'os,  Phys. Lett. B {\bf 331}, 313 (1994);
K. Benakli, Phys. Rev. D {\bf 60}, 104002 (1999); \pl B {\bf 447}, 51 (1999).


\bibitem{quevedo}
C. P. Burgess, L. E. Ibanez and F. Quevedo, Phys. Lett. B {\bf 447},
257 (1999). 

\bibitem{exp1}
For  experimental bounds see for instance:
T. G. Rizzo, Phys. Rev. D {\bf 59}, 115010 (1999);
G. F. Giudice, R. Rattazzi and J. D. Wells,
Nucl. Phys. B {\bf 544}, 3 (1999);
E. A. Mirabelli, M. Perelstein and M. E. Peskin,
Phys. Rev. Lett. {\bf 82}, 2236 (1999);
J. L. Hewett,  Phys. Rev. Lett. {\bf 82}, 4765 (1999);
V. Barger, T. Han, C. Kao and R. J. Zhang, \pl B {\bf 461}, 34 (1999).

\bibitem{exp2}
C. D. Hoyle {\it et al}, Phys. Rev. Lett. {\bf 86}, 1418 (2001).

\bibitem{abdel100}
For a  review see for instance: A. P\'erez-Lorenzana, hep-ph/0008333, to appear in the proceedings of `The IX Mexican School on Particles and Fields', Metepec, Puebla, Mexico, August 2000. 

\bibitem{abdel2} 
A. Mazumdar and A. P\'erez-Lorenzana, Phys. Lett. B {\bf 508}, 340 (2001).

\bibitem{davidson0}
K. Benakli and S. Davidson, \prd {\bf 60}, 025004 (1999);
L. J. Hall and  D. R. Smith. Phys. Rev. D {\bf 60}, 085008 (1999).
M. Fairbairn, Phys. Lett. B {\bf 508}, 335 (2001),
M. Fairbairn, and  L. M Griffiths, hep-ph/0111435.

\bibitem{new}
S. Hannestad, G. G. Raffelt, Phys. Rev. Lett. {\bf 88} 071301 (2002). 

\bibitem{aemp} 
R. Allahverdi, K. Enqvist, A. Mazumdar and A. P\'erez-Lorenzana,
Nucl. Phys. B {\bf 618}, 277 (2001).


\bibitem{abdel3}
A. Mazumdar and A. P\'erez-Lorenzana, hep-ph/0103215.

\bibitem{many}
D. Lyth, Phys. Lett. B {\bf 448}, 191 (1999); 
G. R. Dvali, S. H. H. Tye, Phys. Lett. B {\bf 450}, 72 (1999);
N. Kaloper and A. Linde, Phys. Rev. D {\bf 59}, 101303 (1999);
N. Arkani-Hamed, {\it et al.}, Nucl. Phys. B {\bf 567}, 189 (2000).


 
\bibitem{anu}A. Mazumdar, Phys. Lett. B {\bf 469}, 55 (1999).

\bibitem{abdel1} 
R. N. Mohapatra, A. P\'erez-Lorenzana and C. A. de S. Pires,
Phys. Rev. D {\bf  62}, 105030 (2000).



\bibitem{linde}A. D. Linde, Phys. Lett. B {\bf 259}, 38 (1991); 
Phys. Rev. D {\bf 49}, 748 (1994).

\bibitem{randall}L. Randall, M. Solja\u{c}i\'{c}, and A. H. Guth, Nucl. 
                 Phys. B {\bf 472}, 377 (1996).

\bibitem{garcia} J. Garc\'{\i}a-Bellido, A. D. Linde, and D. Wands, 
                 Phys. Rev. D {\bf 54}, 6040 (1996).




\bibitem{berkin}
A. L. Berkin and K. Maeda, Phys. Rev. D. {\bf 44}, 1691 (1991).


\bibitem{steinhardt}F. S. Accetta, and P. J. Steinhardt, Phys. Rev. Lett.
{\bf 67}, 298 (1991)

\bibitem{bd}C. M. Will, {\em Theory and Experimet in Gravitational Physics},
Cambridge University Press, Cambridge, (1981)


  
\bibitem{fp} A. A. Starobinsky, in {\em Current topics in field theory, 
             quantum gravity and strings}, Lecture Notes in Physics,
             Eds. H. J. de Vega and N. Sanchez (Springer, Heidelberg 1986) 
             {\bf 206}, 107; A. S. Goncharov, A. D. Linde and V. F. Mukhanov,
             Int. J. Mod. Phys. {\bf A2} 561 (1987); A. D. Linde and 
             A. Mezhlumian, Phys. Lett. B {\bf 307}, 25 (1993);
             A. D. Linde, D. A. Linde and
             A. Mezhlumian, Phys. Rev. D {\bf 49} 1783 (1994).


         
\bibitem{bunn} E. F. Bunn, D. Scott and M. White, Ap. J. {\bf 441}, L9 (1995).

\bibitem{msr} M. Sasaki and E. D. Stewart, Prog. Theor. Phys. {\bf 95}, 71 (1996)
\bibitem{pbhed} S. B. Giddings and S. Thomas, Phys. Rev. D {\bf 65},
               056010 (2002); S. Dimopoulos and G. Landsberg, Phys. 
               Rev. Lett. {\bf 87}, 161602 (2001); J. L. Feng and 
               A. D. Shapere, Phys. Rev. D {\bf 88}, 021303 (2002).
\bibitem{pbhedevap} P. C. Argyres, S. Dimopoulos and J. March-Russell,
               Phys. Lett. B {\bf 441}, 96 (1998);
              R. Emparan, G. T. Horowitz and R. C. Myers, 
              Phys. Rev. Lett. {\bf 85}, 499 (2000).
\bibitem{pbh} B. J. Carr, Astrophys. J. {\bf 201}, 1 (1975); B. J. Carr, 
              J. H. Gilbert and J. E. Lidsey, Phys. Rev. D 
             {\bf 50}, 4853 (1994); A. M. Green and A. R. Liddle, Phys. Rev.
              D {\bf 56}, 6166 (1997).

\bibitem{schmaltz}
N. Arkani-Hamed, Y. Grossman, and  M. Schmaltz, Phys. Rev. D {\bf 61},
115004 (2000).

\bibitem{rula}
A. Masiero, M. Peloso, L. Sorbo, and R. Tabbash, Phys. Rev. D {\bf 62},
063515 (2000);  T. Dent, hep-ph/0110318; D. Chung, and T. Dent, hep-ph/0112360. 

\end{references}
\end{document}